# Regenerative Soot-VI: The excited states of neutral and ionized carbon in regenerative sooting discharges

**M Uzair[1], Sajid Hussain[1], S A Janjua[1] and Shoaib Ahmad[1,2,*]**

[1]*Accelerator Lab., PINSTECH, PO Nilore, Islamabad, Pakistan*

[2]*National Centre for Physics, Quaid-i-Azam University Campus, Shahdara Valley, Islamabad, 44000, Pakistan*

[*]Email: sahmad.ncp@gmail.com

**Abstract**

We report the mechanisms of production and the state of excitation of the neutral and singly charged monatomic carbon in the regenerative soot as a function of the discharge parameters in graphite hollow cathode (HC) sources. Two distinctly different source configurations have been investigated. Comparisons of the level densities of various charged states of $C_1$ have identified the regenerative properties of the C radicals in graphite HC soot.

## 1. Introduction

This paper is part of an ongoing activity in which we are investigating the role, that various discharge parameters play, in the formation and later regeneration of the soot in carbonaceous environment created in graphite hollow cathodes (HC). The initial, pure noble gas glow discharge gradually acquires carbonaceous character after kinetic sputtering of the inner walls of the graphite HC with energetic positive ions with K.E (ions) = $q \cdot V_{dis}$ ∼ 500 –1000 eV, where $q$ is the ionic charge. The sputtering of sp$^2$-bonded carbon atoms in graphite leads to the introduction of free carbon atoms and their ions as active ingredients of the discharge, besides the noble gas (He, Ne, etc) atoms and ions $He^{*,+}$, $Ne^{*,+}$ which provide the original support gas plasma. Once created, the $C^{*,\pm}$ participate in the collisional, radiative and capture activities of the multi-component plasma. Formation of carbon clusters $C_x$ ($x \geq 2$) take place in the plasma as well as on the cathode walls, where all neutral species are deposited. The C clusters produce a sooted graphite surface which takes over from the underlying graphite as the effective HC. All subsequent emissions of electrons and the soot constituents from the cathode are in fact from this sooted layer. This process has similarities with the soot production by arc discharge of Kratschmer *et al* [1], however, the main difference is in the continuous operation of the sooted source. In our case this recycles the soot from which C atoms, ions and clusters are emitted into the discharge. This process is called the regeneration of the soot by which one can manipulate the overall constitution of the soot by adjusting the discharge parameters. The regenerative sooting environment that leads to the formation of $C_x$ ($2 \leq x \leq 10\,000$) has been studied with the twin techniques of mass spectrometry and emission spectroscopy; a detailed study has recently been published else-where [2]. These studies are done with our specially designed sources with graphite HC operating in cusp magnetic fields. Depending upon the state of sooting of the cathode and the discharge parameters, we have been able to extract from these sooting discharges, a broad range of carbon clusters and studied the VUV emissions from neutral (C I), singly charged (C II) and the doubly charged (C III) monatomic carbon [3]. We produce and subsequently regenerate the soot in these sources whereas, another related series of experiments were conducted in a non-regenerative environment that highlighted the negative carbon atoms and clusters ($C_1^-$, $C_2^-$, $C_3^-$ and $C_4^-$), as the dominant species emitted from ion bombarded graphite [4].

Regeneration of the soot in graphite HC is therefore, a useful mechanism to study the processes of formation and fragmentation of the soot. It can also be used to investigate the monatomic carbon emission lines because $C_1$ is an essential by-product of the C cluster fragmentation sequences [5]. We have investigated and presented here: (a) the effect of two different source geometries on the state of excitation of C I and C II, (b) the variations of $C_1$ component of the carbonaceous vapour as a function of the cycling of the discharge current $i_{dis}$ and (c) the conspicuous role of the singly charged C II revealed by its first excited metastable level density through the $\lambda$ =233 nm inter-combination (IC) multiplet ($^2P_{1/2,3/2}$–$^4P_{1/2,3/2,5/2}$). Another of the CII IC singlet at 220 nm is also a major component of these discharges that yields large densities of the upper quartet levels

## 2. Sooting discharges
### 2.1. Experimental

The carbon vapour in these sooting discharges is produced in two coaxial, inter-penetrating graphite hollow cylinders. By alternating the inner and the outer graphite cylinders that operate as HC and anode (A), respectively, we investigate the effect of the source geometry on the processes of a unique carbon vapour whose constitution and properties can be manipulated. The inset in figure 1 shows the geometry of the source with a cylindrical graphite hollow anode (HA) (25 mm long with inner and outer diameters being 14 mm and 16 mm, respectively) surrounded by 18 mm inner diameter HC of 54 mm length. The two cylinders can, however, be interchanged and one can have the traditional, well studied central cylinder as HC [6]. Charged $C_1$ and C clusters $C_x$ ($x \geq 2$) are extracted by biasing the source with respect to the extractor while a fused silica window lets a small amount of the light out, for emission spectroscopy. All emissions are along the axial direction and represent plasma activity along the HA–HC axis. The emission spectrum in the figure is obtained during the initial stages of soot formation with Ne at 0.1 mbar, $i_{dis}$=75 mA at discharge voltage $V_{dis}$=0.6 kV with a Jobin Yvon monochromator with a grating blazed at 300 nm with 1 Å resolution. Data acquisition with moderate stepper motor step size take about 20 min for 180–510 nm range. A discharge stabilization time of 15 minutes is given between successive parametric changes. Pressure increase is observed during the initial sputtering dominated stages but it stabilizes when the source is operating in the sooting mode. The quantum efficiency of the photomultiplier tube and the relative efficiency of the grating in the range of our measurements 185–600 nm are taken into account while converting the respective experimentally observed line intensities $I_{obs}$ into the calibrated energy emitted as photons per sec per unit solid angle d per unit volume of the source–$I_0$. This yields us the calculated line intensity $I_{mn}$ of the transition between energy levels $m$ and $n$ as $I_0 4\pi \, (d)^{-1}$ (eV s$^{-1}$ cm$^{-3}$). The number density of the upper level in a given volume

$$N_m = I_{mn}/(h\nu_{mn} \cdot A_{mn})$$

where $h\nu_{mn} = E_m - E_n$ (eV) is the energy difference between the levels $m$ and $n$ and $A_{mn}$ (s$^{-1}$) its Einstein transition probability for spontaneous emission. The formula for the evaluation of the level density from the observed line intensity is valid for all spontaneous emissions whether the excited species is in thermodynamic equilibrium or not [7].

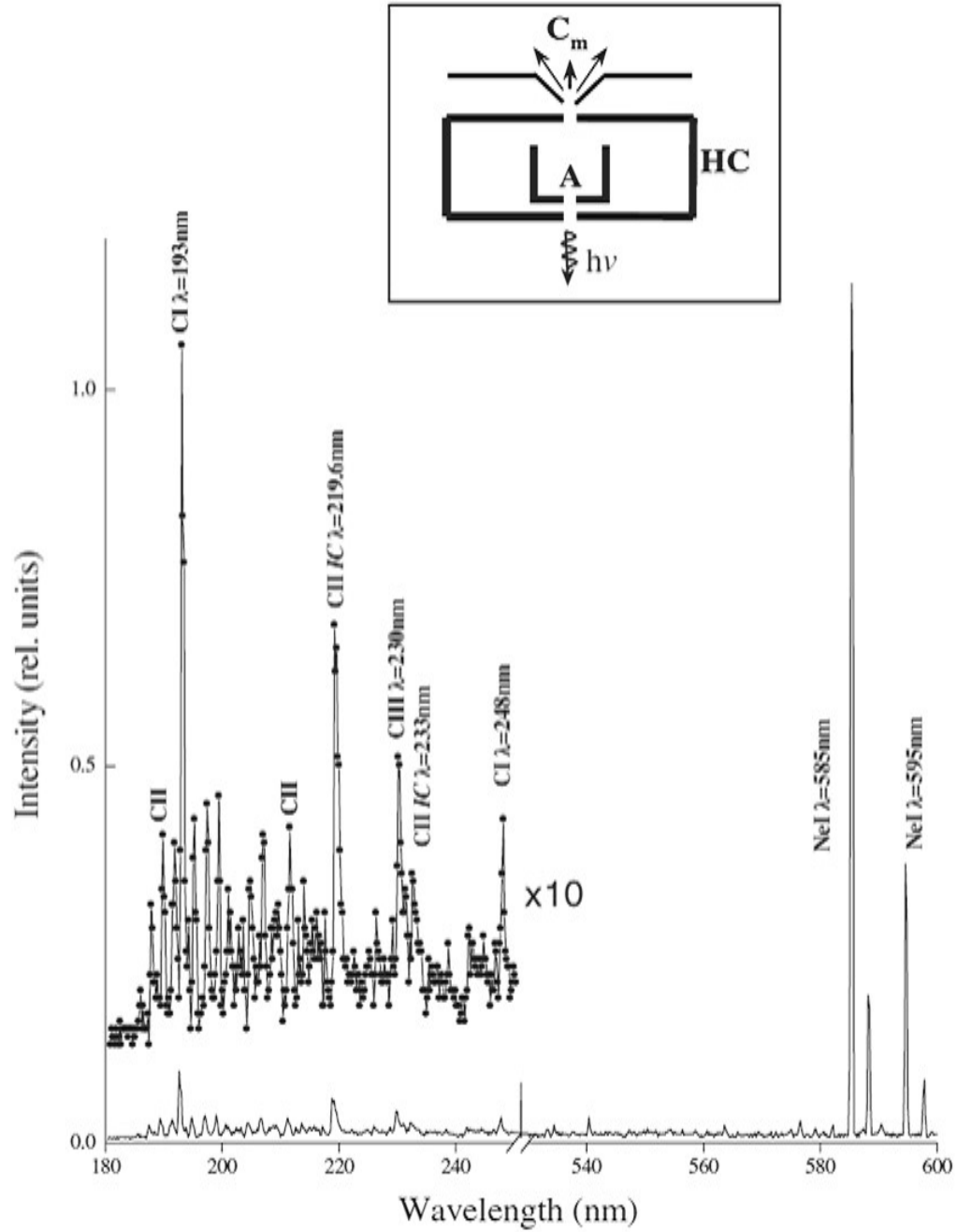

Figure 1. The emission spectrum of the Ne discharge is shown from graphite hollow cathode in the initial stages of source operation with the emission lines from the excited and ionized monatomic carbon's C I, C II and C III lines whose 10 times enlarged intensity between 180 and 250 nm identify typical lines from a mildly sooting plasma. Large C I peaks are flanked with smaller C II and C III emission lines. Inset shows the schematics of the hollow cathode source with two graphite hollow cylinders. The source can emit carbon clusters $C_m$ and the light escapes through fused silica window and the emission lines from plasma species neutral as well as charged can be observed.

## 2.2. Initiation of the carbonaceous discharge

In figure 1 Ne I emission lines are identifiable between $\lambda$ =540 and 600 nm that decay to the first excited states. That indicates a large fraction of the metastable Ne atoms which act as efficient potential sputtering agents for the regeneration of the soot [8]. Two excited atomic lines of C I at $\lambda$ =193.1 nm ($^1$D–$^1$P) and $\lambda$ =247.8 nm ($^1$S–$^1$P) are the signatures of the cathode sputtering and for the generation of a carbonaceous discharge. These two C I emission lines can also act as markers and are used for the absolute calibration of the spectra. The graphite HC discharge with Ne as the source gas, shows two distinct groups of emission lines between 180 and 600 nm. The first group between 180 and 250 nm includes emission lines of C I, C II, C III and C IV. The second distinct and high intensity group of Ne I emission lines lies between 580 and 600 nm. A 10 times enlarged view of 180–250 nm is also shown to highlight that,

(a) the source has just begun to sputter the maiden graphite surface with Ne ions while later on the positively charged C ions and charged clusters also participate in kinetic sputtering of cathode,
(b) the C content of the discharge has a sizable monatomic population and
(c) the positively charged $C_1$ is an essential component of the discharge in which Ne or He are the support gases. Our experiments have shown and we will prove here that different stages of soot formation and subsequent regeneration can yield varying amounts of C I and C II as a function of the discharge parameters and the state of sooting. This implies that one can recycle or regenerate the soot that has already been formed.

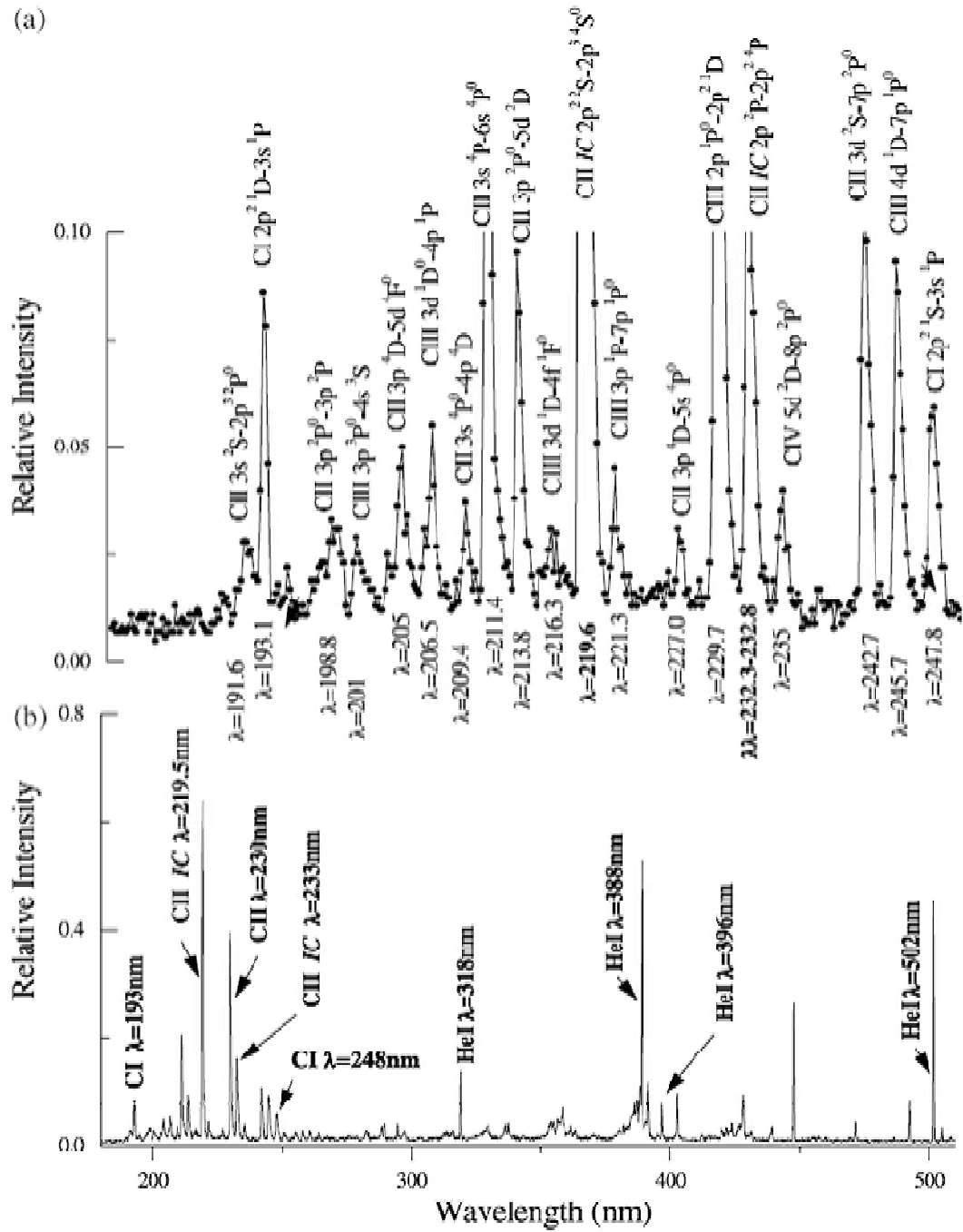

Figure 2. The relative lines intensities in (a) and (b) show the discharge species occupying two wavelength ranges; the emissions lines belonging to C I, C II and C III are prominent within $\lambda = 180$–250 nm and the four labelled transitions of He I between 318 and 502 nm. This emission spectrum is from a well sooted source. In (b) He I lines dominate the higher wavelength range while the C lines are of similar intensity. A 10 times intensity enhanced view of $\lambda = 180$–250 nm range is shown in (a) to identify the transitions. In (b) the C II and C III emission lines are the dominant species of the plasma.

## 2.3. The sooting discharge and recycling of C I and C II

Figure 2 shows that the graphite HC discharge with He. The spectrum is taken after the source has been operated in Ne for about 10 h at 0.1 mbar. This prior treatment is necessary, as the source cannot be operated at low pressures with He in an un-sooted HC. Therefore, we are operating in an environment where soot has already been formed and deposited on the cathode walls. Similar to the spectrum with Ne as the source gas, figure 2 also shows two well-defined groups of emission lines between 180 and 510 nm; the first is the group of UV lines between 180 and 250 nm in figure 2(*a*) belonging, as in figure 1, exclusively to C I, C II and C III. The second distinct and high intensity group of He I emission lines lies between 300 and 510 nm. Figure 2(*b*) shows the complete emission spectrum from the sooting discharge with He as the support gas. Helium pressure is maintained at 2 mbar and the source is operated at discharge current $i_{dis}$ = 25 mA. The characteristic line emission in the range 180–250 nm can be monitored as a function of discharge parameters. The intensity scale has been enlarged in figure 2(*a*) to identify the large number of emission lines due to the highly excited atomic and ionic levels of $C_1$. We can clearly identify emission lines of which 2 are due to C I, 8 of the allowed transitions and 2 IC multiplets of C II, 7 transitions belong to C III and 1 multiplet due to C IV (5d $^2$D–8p $^2$P$^0$ ). The transitions have been identified using the NIST Atomic Spectra Database [9]. We have plotted the data points to emphasize the reproducibility of these peaks even though the resolution of our monochromator is 0.1 nm. The comparison of emissions in figures 1 and 2(*a*) shows that the ratio of C I to C II and C III peaks reduce as we go from a mildly sooted source operated with Ne to a well-sooted one with He. The C II IC lines are significantly enhanced in figure 2(*a*). Helium's excited lines are also pointed out in figure 2(*b*) especially the transitions at $\lambda$ = 318.7, 388.8, 395.6 and 510.5 nm. These lines identify the population densities of the metastable levels 2s $^3$S and 2s $^1$S, respectively, that are being produced as a result of these transitions. These emission lines are also used to calculate the excitation temperatures by using the *S* → *S* and *T* → *T* lines of He. A multi-component glow discharge whose composition is being recycled may not be in local thermodynamic equilibrium (LTE). The ongoing processes of the soot regeneration, that involve formative as well as fragmentation stages of the clusters, continuously modify the discharge composition. From such a non-LTE plasma, one can only calculate the excitation temperatures $T_{exc}$ of different species rather than the typical electron temperature $T_e$ for the plasma. The ratios of the upper $N_u$ and lower $N_l$ level densities for respective transitions to the same level *k* are used for calculation of $T_{exc}$. The line intensities $I_{uk} = N_u \cdot A_{uk} \cdot h\nu_{uk}$ and a similar one for $I_{lk}$ are of the transitions between the three levels *u*, *l* and *k*. These are used for the evaluation of the relative ratio of level densities $N_u/N_l$. The third level is the common one, and the ratio of the two level densities is given by Sobel'man [7] as $N_u/N_l = (g_u/g_l) \exp\{-(E_u - E_l)/kT_{exc}\}$, where $g_u$, $g_l$ are the statistical weights and $h\nu_{ul} = E_u - E_l$ , the energies of the respective transitions. For our experiments $T_{exc}$(He)~0.35–0.5 eV.

While evaluating the pattern of $C_1^-$ emission lines, we place a special emphasis on the 233 nm IC multiplet. It is the spin-forbidden transition $^2P_{1/2,3/2}$–$^4P_{1/2,3/2,5/2}$ with a cumulative transition probability $\Sigma A \simeq 215\ s^{-1}$. While the individual transition probabilities are reported in ref [9]. Both the theoretical and experimental studies of various researchers including atomic and astrophysicists [10] have indicated oscillator strengths and transition probabilities within 10% of the values shown in the NIST values for $^2P_{1/2,3/2}$–$^4P_{1/2,3/2,5/2}$ transitions. In our spectra the 233 nm peak is dominated by the two transitions at

$\lambda$ = 232.47 nm, $J$ =1/2–1/2 , $\tau$ = 13.7 ms and at $\lambda$ = 232.54 nm, $J$ =3/2 –5/2 , $\tau$ = 19 ms. The multiplet is peaked around 232.5 ± 0.1 nm. These long-lived metastable ions of the singly charged carbon are used to identify and determine the mechanisms of electron induced collisions.

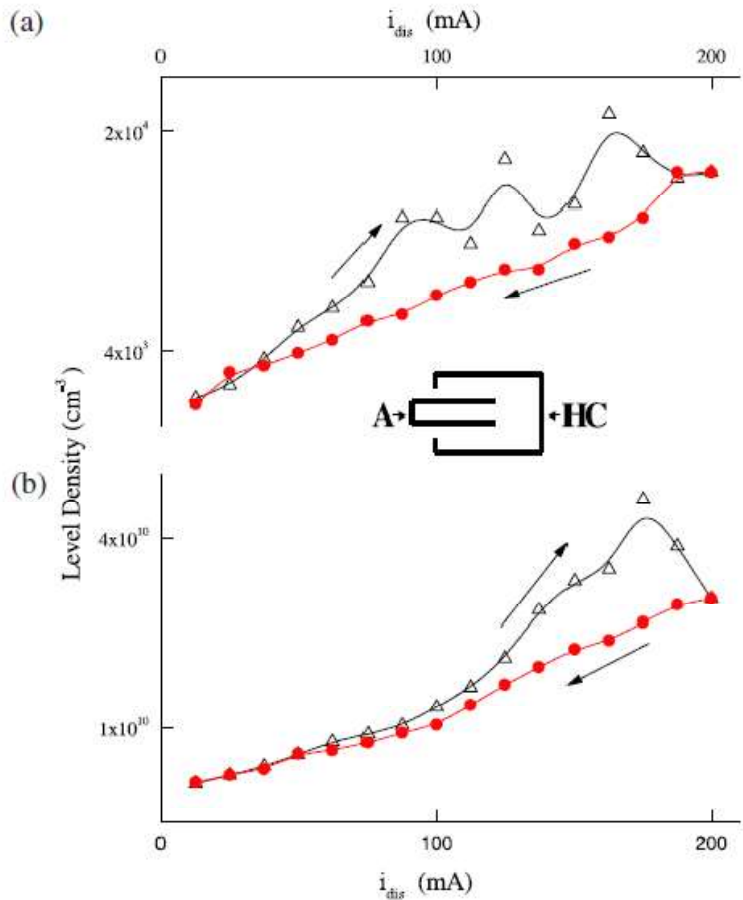

**Figure 3.** The level densities of C I's $^1P$ level from $\lambda = 248$ nm ($^1S$–$^1P$) and C II's first excited level $^4P_{1/2,3/2,5/2}$ from the intensities of the IC multiplet at 233 nm are presented in (*a*) and (*b*), respectively, as a function of $i_{dis}$. The geometrical arrangements of the hollow cathode is also shown. The discharge current is varied from 12.5 to 200 mA and then reduced back to 12.5 mA; the cycling sequence being 12.5→200→12.5 mA.

Upper level densities of C I and C II transitions are plotted in figure 3 from the measured intensities of C I $\lambda$ = 247.8 nm ($^1S$–$^1P$) and C II IC multiplet at 233 nm as a function of $i_{dis}$. The source is operated in the increasing and decreasing cycles of $i_{dis}$ while monitoring the emission spectra. The parameter for the regeneration of the soot is $i_{dis}$ that has been varied between 12.5 and 200 mA with a period of 10 min between each change to allow the discharge to stabilize. In this process the overall discharge is regenerated with the populations of different excitation and ionization states establishing a net equilibrium at the ambient discharge conditions. The source configurations is shown in the inset of figure 3(*a*) with central anode (A). In figure 3(*a*) the pattern of increase of 3s $^1$P (CI) level density is such that it rises to ∼$10^4$ cm$^{-3}$ at $i_{dis}$ = 75 mA and show fluctuations in the range of $i_{dis}$ =75–200 mA. From 200 to 12.5 mA, the level density monotonically decreases to its equilibrium value at 12.5 mA. Figure 3(*b*) shows the metastable level density of C II $^4P_{1/2,3/2,5/2}$ that is responsible for the 233 nm multiplet. Its density gradually increases with $i_{dis}$ up to 100 mA followed by a sharp increase in the range 100–175 mA and then decreases by a factor of 2 between 175 and 200 mA. The decreasing pattern also has two slopes, a sharper one between 200 and 100 mA and a slower one between 100 and 12.5 mA.

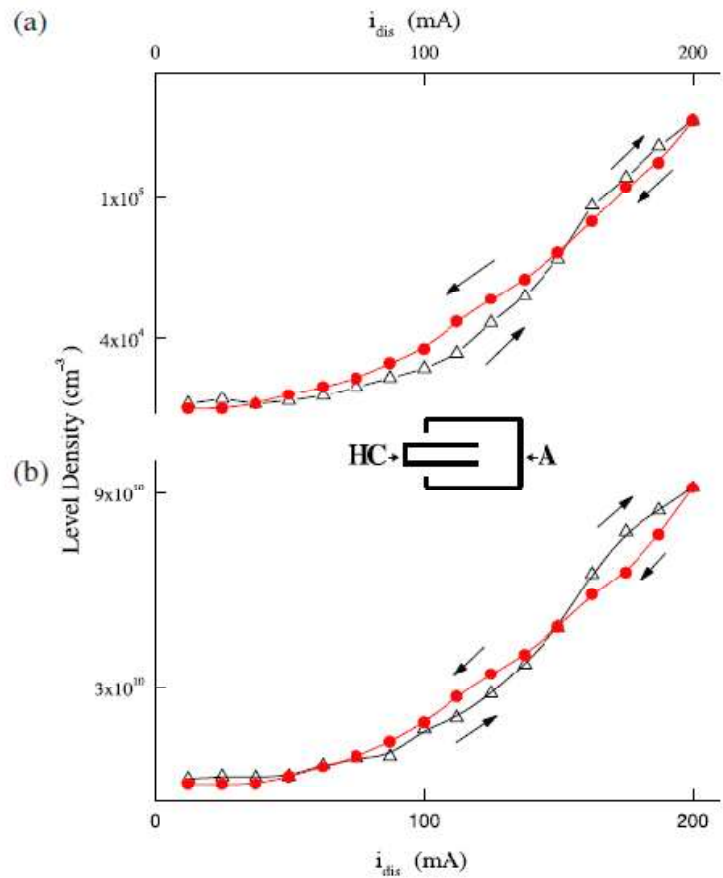

Figure 4. For an inverted source configuration the same level densities of C I and C II as in figure 3, are plotted against $i_{dis}$ in the same discharge current range.

In figure 4(*a*) we have the conventional HC geometry with the inner graphite hollow cylinder being the cathode and the outer one serving as the anode. The respective level densities are about an order of magnitude higher in this configuration compared with those in figure 3. Level densities of C I ($^1$P) and C II ($^4P_{1/2,3/2,5/2}$) are plotted in figures 4(*a*) and (*b*), respectively, and both the species have similar level variation pattern with two regimes; one between 12.5 and 100 mA and the other between 100 and 200 mA. In the higher range of $i_{dis}$, there is a flip-flop between the increasing and decreasing level densities.

## 2.4. The hysteresis in level densities of C I and C II

The relative error in determining the level densities is around 10–12% during the $i_{dis}$ cycling between the lowest and the highest values. As such, the hysteresis observed is an interesting feature of our regenerative sooting discharges as a function of the HC geometry. The hysteresis of level densities of C I and C II in figure 3 as a function of $i_{dis}$ indicates the competition of the mechanisms of collisional excitations between the two energizing species of the discharge, i.e. electrons and the high potential energy metastable atoms and ions. It leads to two competing mechanisms: (a) electron-collisional induced processes and (b) inter-particle collisions. On increasing $i_{dis}$ a large number of excited species are created besides the electrons. The electron density $n_e$ is directly proportional to the dis-charge current $i_{dis}$ whereas the excited and ionized species have a higher dependence on $i_{dis}$ and has been associated with 2- and 3-body collisions of the discharge species among themselves [6]. The data of the level densities of C I and C II presented in figures 3 and 4 relate to the emissions along the HA–HC axis in the two inverted geometries. In figure 3 a much larger surface area of HC is available for carbon sputtering from the cathode that leads to the excited and ionized monatomic $C_1$. In addition, an order of magnitude larger discharge volume containing the electrons, sputtered atoms, molecules and clusters is available. While in figure 4 a smaller sputtering cathode is available nearer to the HA–HC axis. Therefore, the level densities are 5–10 times higher in figure 4 for the corresponding species in figure 3. However, the actual level densities show a hysteresis only in the case of a larger and outer HC in figure 3. The collisions between the excited, sputtered species may be responsible for up to 30% more C I and C II during the increasing cycle of $i_{dis}$ in figure 3 than in figure 4.

## 3. The excitation and ionization mechanisms of $C_1$

An energy level diagram of C I and C II is shown in figure 5. Only those levels that result in the emission of some of the observed transitions of C I and C II monatomic C are shown. While only two of C I emission lines are shown, the main emphasis is on describing the state of excitation of C II via radiative and the autoionizing (AI) transitions. The D→D transitions are present in all of the spectra but mostly from those levels that are below the AI level. On the other hand, the transitions from the quartet levels have been seen originating both from the AI levels and those levels that are below the C II ionizing limit. The diagram shows the typical number densities of the respective levels. To interpret the results of the study of the highly excited C II emissions from the regenerative sooting discharge, we have monitored the pattern of increase and decrease of the unusually intense 233 nm IC multiplet of C II. In figure 2(*a*) 7 C III transitions were identified; 6 of these are S→S transitions and one among the triplet states. One C IV transition was also identified occurring between the highly excited levels. As can be seen from the labelling of transitions in figure 2(*a*), there are a number of quartet–quartet (Q→Q) transitions of C II from the AI levels. Three such transitions occur from within the AI levels; at $\lambda$ = 205, 211.4 and 227 nm. The ratio of the AI transition probability ($A_{AI}$) to that of spontaneous radiative transition ($A_R$) from the same AI level for producing the doubly charged carbon C III in its ground state $^1S$ is $A_{AI}/A_R \sim 10^5$ as discussed by Bely and Regemorter and the references therein [11]. Therefore, these observed radiative transitions of C II indicate a large population density of C III in the regenerative sooting discharge. Its magnitude can be estimated by using a typical Q→Q, C II radiative transition yields $A_R \sim 10^5$ cm$^{-3}$ which implies that C III ground state ion density $A_{AI} \sim A_R \cdot 10^5 \sim 10^{10}$ cm$^{-3}$. From the energy level scheme and using NIST's Atomic Database [9], we know that a total of 14 C II multiplets are emitted by de-excitation of the quartet terms to $^4P_{1/2,3/2,5/2}$ with a cumulative transition probability $A_i$ = 432 × 10$^8$ s. This in turn de-excites to the ground term $^2P_{1/2,3/2}$ by the emission of the 233 nm multiplet. If we compare this with the similar number for the doublet-doublet transitions then $A_i$ = 249 × 10$^8$ s. The intense emission indicates that C II exists as a highly excited C ion in the discharge. The cumulative decay scheme of the highly excited C II is shown in figure 5 which suggests that the Q→Q transitions are the dominant emission processes that eventually populates the $^4P_{1/2,3/2,5/2}$ metastable level.

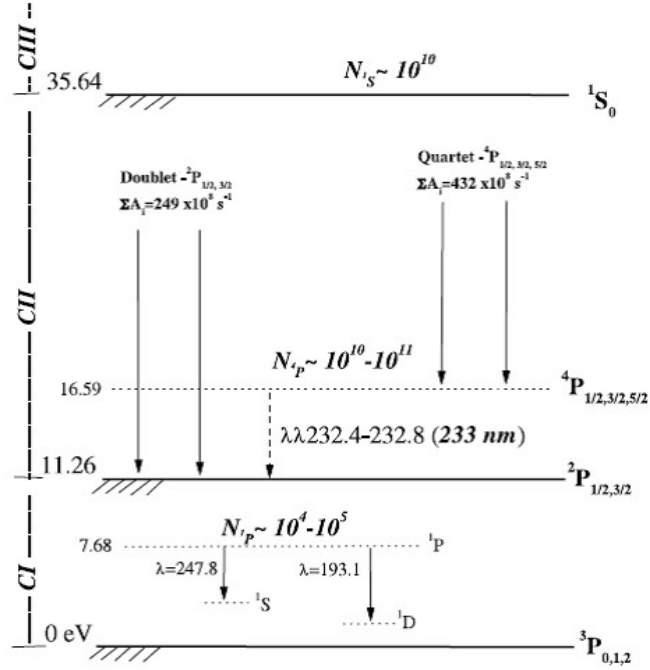

**Figure 5.** The energy levels diagram of C I and C II shows some of the representative transitions and the calculated number densities of the respective levels. The allowed as well as the intercombination lines are indicated. Unbroken arrows show the allowed and the broken lined arrows are for the intercombination transitions. The ground states of C I ($^3P_{0,1,2}$), C II ($^2P_{1/2,3/2}$) and C III ($^1S_0$) are indicated.

The most significant aspect of such discharges is that these are initiated and sustained by two well-defined electron energy regimes with Maxwell velocity distributions [6]; one with energies 10 eV and the other with 1 eV. The high energy electrons ( 10 eV) are emitted from the graphite cathode and accelerated by the transverse electric field in the cathode dark space, can efficiently ionize. These electrons ionize the support gas as well as other species which in turn sputter the cathode. Lotz's [12] ionization rate coefficient $\alpha$ is used to estimate the relative number densities of the successively higher C ions. At electron energy of 1 eV that corresponds to the excitation temperature $T_{exc}$ the corresponding ionization rates

$\sim 10^{-13}$ cm$^3$ s$^{-1}$, $\alpha_{C\,II \to C\,III} \sim 10^{-19}$ cm$^3$ s$^{-1}$ and

$\sim 10^{-30}$ cm$^3$ s$^{-1}$.

At these electron temperatures one cannot expect to see the level of ionization that we have shown in figures 1 and 2. While at 10 eV the corresponding rates are enhanced by many orders of magnitudes as

$\alpha_{C\,I \to C\,II} \sim 10^{-8}$ cm$^3$ s$^{-1}$,

$\alpha_{C\,II \to C\,III} \sim 10^{-9}$ cm$^3$ s$^{-1}$ and

$\alpha_{C\,III \to C\,IV} \sim 10^{-11}$ cm$^3$ s$^{-1}$.

These high energy electrons ( 10 eV) ensure high level densities of the multiply charged carbon $C^{q+}$ ($q$ 1). The radiative processes of C II seem to identify that the energy transferred to the plasma species from the high energy electrons efficiently populates ionized C.

## 4. Conclusions

In conclusion, the multi-component sooting discharge indicates that radiative cascading populates the first excited state of the singly charged carbon. Only this explanation can provide justification to the 4–5 orders of magnitude more intense $^4P_{1/2,3/2,5/2}$ level densities compared with those of the doublets and the quartets. This may be possible only if $C_1$ exists in a highly ionized state and is radiatively recombining from C III to C II. The C II IC line at $\lambda$ =219.6 nm is the other unusual feature of all of our spectra with Ne and He at all discharge voltages, currents and gas pressures. It is a quartet→doublet spin-forbidden transition $2p^2\ ^2S_{1/2}$-$2p^3\ ^4S_{3/2}$ with probability $A$ 5. 1 s$^{-1}$ · $2p^3\ ^4S_{3/2}$ level has high density ∼$10^{11}$ cm$^{-3}$. We speculate that resonant electron collisional processes may be responsible for the high population density of this particular level. It is pertinent to point to the fact that at 2 mbar pressure of He inter-particle collisions have a shorter time than the C II metastables. This may be a competing mechanism for the charge transfer from C II to He and vice versa. The collisions between the excited, sputtered species may be responsible for the hysteresis effect seen in the level densities of C I and C II during the increasing cycle of $i_{dis}$ in figure 3.

As a multi-component plasma, our regenerative soot has variable characteristics for the different species in the discharge. It is this unique property of the regenerative soot that makes it ideal to study the state of excitation and ionization of the carbon species that may not be in LTE. One of the objectives of this paper is to identify that the state of excitation and ionization of monatomic C does not follow the electron-collisional excitations dictated by Boltzmann distribution for a plasma in LTE. The carbonaceous discharge in graphite HC can provide valuable new insights into the state of the excitation, ionization and recombinations (radiative as well as dielectronic) of $C^+_1$ that may be exploited for designing VUV sources and systems as has recently been suggested [3].

**References**

[1] Kratschmer W, Fostiropoulous K and Huffman D R 1990 *Chem. Phys. Lett.* **170** 167–70
[2] Ahmad S 2002 *Eur. Phys. J.* D **18** 309–18
[3] Ahmad S 2003 *Eur. Phys. J.* D **22** 189–92
[4] Qayyum A, Akhtar M N, Riffat T and Ahmad S 1999 *Appl. Phys. Lett.* **75** 4100–2
[5] Rangavachari K and Binsky J S 1987 Physics and chemistry of smalll cluster *NATO Advanced Studies Institute Series B:Physics* vol 150, ed P Jena *et al* (New York: Plenum) p 317 ;
Ahmad S, Ahmad B and Riffat T 2001 *Phys. Rev.* E **64** 026408–14
[6] Falk H 1985 *Improved Hollow Cathode Lamps for Atomic Spectroscopy* ed C Sergio (Chichester: Ellis Horowood) chapter 4
[7] Sobel'man I I 1972 *Introduction to the Theory of Atomic Spectra* (Oxford: Pergamon) chapter 9;
Lang K R 1974 *Astrophysical Formulae* (Berlin: Springer) chapter 2
[8] Ahmad S and Akhtar M N 2001 *Appl. Phys. Lett.* **78** 1499–501
[9] NIST Atomic Spectra Database (ADS) Data at http://physics.nist.gov/
[10] Stencel *et al* 1981 *Mon. Not. R. Astron. Soc.* **196** 47P–53P;
Nussbaumer H and Storey P J 1981 *Astron. Astrophys.* **96** 91–5;
Fang *et al* 1993 *Phys. Rev.* A **48** 1114–22
[11] Bely O and Van Regemorter H 1970 *Ann. Rev. Astron. Astrophys.* **8** 329–67; Lotz W 1967 *Astrophys. J.* **14** 207–38